\documentclass[twocolumn]{revtex4}
\usepackage{amssymb}
\usepackage{latexsym}
\usepackage{epsfig}
\usepackage{color}
\usepackage{float}
\usepackage{graphicx}
\usepackage{epstopdf}

\begin{document}

\title{Generalized Misner-Sharp energy in the generalized Rastall theory}
\author{H. Moradpour\footnote{hn.moradpour@maragheh.ac.ir}, M. Valipour}
\address{Research Institute for Astronomy and Astrophysics of Maragha (RIAAM), University of Maragheh, P.O. Box 55136-553, Maragheh, Iran}

\begin{abstract}
Employing the unified first law of thermodynamics and the field
equations of the generalized Rastall theory, we get the
generalized Misner-Sharp mass of spacetimes for which
$g_{tt}=-g^{rr}=-f(r)$. The obtained result differs from those of
the Einstein and Rastall theories. Moreover, using the first law
of thermodynamics, the obtained generalized Misner-Sharp mass and
the field equations, the entropy of the static spherically
symmetric horizons is also addressed in the framework of the
generalized Rastall theory. In addition, by generalizing the study
to the flat FRW universe, the apparent horizon entropy is also
calculated. Considering the effects of applying the Newtonian
limit to the field equations on the coupling coefficients of the
generalized Rastall theory, our study indicates $i$) the obtained
entropy-area relation is the same as that of the Rastall theory,
and $ii$) the Bekenstein entropy is recovered when the generalized
Rastall theory reduces to the Einstein theory. The validity of the
second law of thermodynamics is also investigated in the flat FRW
universe.
\end{abstract}

\maketitle

\section{Introduction}

Based on the curvature-matter non-minimal coupling theories, the
ordinary energy-momentum conservation law is not valid
\cite{cmc,cmc1,cmc2}, a hypothesis that returns to Rastall
\cite{rastall}. In this framework, energy can be exchanged between
the matter source and geometry leading to a modified version of
general relativity \cite{m1} that has interesting outcomes
\cite{m1,m2,m3,m5,m6,m7,m8,m9,m10,m11,m12,m13,m14} and for
example, lets baryonic matter to support traversable wormholes
\cite{m4}. Recently, introducing a generalized version for the
Rastall theory, it has been shown that such non-minimal coupling,
and indeed, the geometry ability to couple with the matter fields
in a non-minimal way, can theoretically describe inflation and the
current accelerated universe without need for considering the dark
energy candidates and inflationary fields \cite{epjc}.

The profound connection between thermodynamics and gravitational
theories
\cite{haw,GSL,GSL0,jacob,jacob1,lt,cons,cons1,cons2,cons3,Cai2,CaiKimt}
motivates physicists to look for the Misner-Sharp mass in the
various gravitational theories for studying the thermodynamic
aspects of the theories \cite{ms,ref0,ref1,ref2,pad,pad1,pad2,pad3}.
The same analysis has been done in the Rastall framework
\cite{rasah} indicating that only when this theory reduces to the
Einstein theory, the Bekenstein entropy is recovered in the static
spherically symmetric spacetimes \cite{rasah}, a result in agreement
with those of the dynamic studies \cite{rasen,plb}.

In Ref.~\cite{nepjc}, studying some cosmological consequences of the
generalized Rastall theory \cite{epjc}, authors address some
similarities between the generalized Rastall theory and other
cosmological frameworks such as the Einstein gravity with particle
creation mechanism \cite{pcp} such that the Rastall parameter is
related to the particle creation parameter. They also used the
Misner-Sharp mass of the Einstein theory \cite{ms} as well as the
Cai-Kim temperature \cite{Cai2} in order to obtain the entropy of
the apparent horizon of FRW spacetime. Based on their results, the
generalized Rastall theory preserves the Bekenstein bound of entropy
\cite{nepjc}.

Here, we are interested in finding the generalized Misner-Sharp mass
in the generalized Rastall theory by using the thermodynamic laws
and the corresponding field equations. Our final goals are also to
find the entropy corresponding to the horizons of the static
spherically symmetric spacetime and the flat FRW universe. We
present our analysis and results in the next section, where the
requirements for the validity of the second law of thermodynamics
will also been addressed. The last section is devoted to a summary.
The unit of $c=\hbar=1$ is used throughout this paper.

%%%%%%%%%%%%%%%%%%%%%%%%%%%%%%%%%%%%%%%%%%%%%%%%%%%%%%%%%%%%%%%%%%%%%
\section{Thermodynamics of the generalized Rastall theory}

Based on the generalized Rastall theory \cite{epjc}

\begin{eqnarray}\label{gr0}
T^{\mu\nu}_{\ \ \ ;\mu}=(\lambda R)^{;\nu},
\end{eqnarray}

\noindent which finally leads to

\begin{eqnarray}\label{gr}
G_{\mu\nu}+\kappa\lambda g_{\mu\nu}R=\kappa T_{\mu \nu}.
\end{eqnarray}

\noindent where $\kappa$ is an unknown constant, called the Rastall
gravitational coupling constant, and $\lambda$ is the Rastall
parameter. The field equations look very similar to those of the
Rastall theory, and indeed, only one difference exists. Unlike the Rastall
theory, the Rastall parameter is variable here. Applying the
Newtonian limit to this field equations, one easily reaches

\begin{eqnarray}\label{k2}
\frac{\kappa}{4\kappa\lambda-1}(3\kappa\lambda-\frac{1}{2})=\kappa_G,
\end{eqnarray}

\noindent in which $\kappa_G\equiv4\pi G$, meaning that this
generalized Rastall theory lets $G$ change, because $\lambda$ is
not generally constant. As a result, one easily obtains that if we
presume $\kappa\equiv8\pi G$, then the Newtonian limit
automatically indicates that $\lambda=0$ meaning that we are in
the Einstein framework, and thus the Einstein results should be
recovered. Eq.~(\ref{k2}) is similar to that of the Rastall theory
\cite{rastall,rasah}, a result due to the fact that both theories
include the same additional geometrical term (the Ricci scalar)
compared with the Einstein theory.

%Finally, by defining $\eta=\kappa\lambda$, Eq.~(\ref{k2}) yields
%
%\begin{eqnarray}\label{eta}
%\kappa=\frac{4\eta-1}{6\eta-1}2\kappa_G,\ \
%\lambda=\frac{\eta(6\eta-1)}{(4\eta-1)2\kappa_G},
%\end{eqnarray}

%\noindent implies $\lambda=0$ is equal to $\eta=0$.
The unified first law of thermodynamics is written as \cite{ref1,ref2}

\begin{eqnarray}\label{ufl}
dE\equiv A\Psi_a dx^a + W dV.
\end{eqnarray}

\noindent Here, $\Psi_a=T^b_a\partial_b r + W\partial_a r$ and
$W=-\frac{h^{ab}T_{ab}}{2}$ denote the energy supply vector and the
work density, respectively, where $h^{ab}$ is the metric on the
hypersurface ($t,r$). Additionally, $A=4\pi r^2$ is the area of the
system boundary located at radius $r$. For the FRW and static
spherically symmetric spacetimes, the apparent and event horizons
are proper causal boundary, respectively
\cite{haw,GSL,GSL0,jacob,jacob1,lt,ms,ref0,ref1,ref2,pad,pad1,pad2,pad3,rasah,cons,cons1,cons2,cons3,Cai2,CaiKimt,nepjc,rasen,plb}.

%%%%%%%%%%%%%%%%%%%%%%%%%%%%%%%%%%%%%%%%%%%%%%%%%%%%%%%%%%%%%%%%%%%%%%%%%%%%%%%%%
\subsection{Thermodynamic analysis of the static spherically symmetric spacetimes}

Now, let us follow Refs.~\cite{pad,rasah} and consider the
spherically symmetric static spacetime

\begin{equation}\label{met2}
ds^{2}=-f(r)dt^{2}+\frac{dr^{2}}{f(r)}+r^{2}d\Omega^{2},
\end{equation}

\noindent while it is supported by an energy-momentum source with
energy density $\rho$ and pressure $p$ (or equally the
energy-momentum tensor $T^\mu_\nu=diag(-\rho,p,p,p)$), and its
horizon is located at $r_h$ (leading to $f(r_h)=0$). Indeed, field
equations let us to interpret $T^\mu_\nu$ as the energy-momentum
tensor produced (induced into the spacetime) by the (existence of)
black hole with radius $r_h$. In this manner, Eq.~(\ref{ufl})
easily leads to

\begin{eqnarray}\label{ufl10}
dE=4\pi r^2\rho dr.
\end{eqnarray}

\noindent Using the $t-t$ component of~(\ref{gr}), one can get
$\rho$ placed in the above equation to reach at

\begin{eqnarray}\label{E1}
&&dE=\frac{4\pi}{\kappa}\big[1-\frac{d(rf(r))}{dr}+\\&&\kappa\lambda[\frac{d(r^2f^{\prime}(r))}{dr}-2(1-\frac{d(rf(r))}{dr})]\big]dr.\nonumber
\end{eqnarray}

\noindent This is in fact the differential of the Misner-Sharp mass
content of the generalized Rastall theory introduced in \cite{epjc},
and its integral leads to

\begin{eqnarray}\label{E2}
&&E=\frac{4\pi}{\kappa}\big[r\big(1-f(r)\big)+\\&&\kappa\int\lambda[\frac{d(r^2f^{\prime}(r))}{dr}-2(1-\frac{d(rf(r))}{dr})]\big]dr.\nonumber
\end{eqnarray}

\noindent In general, one can find $E$ by knowing the exact from of
$\lambda$. For the Misner-Sharp mass confined to the radius $r_h$,
this equation leads to

\begin{eqnarray}\label{E21}
&&E_h=\frac{4\pi}{\kappa}\bigg[r_h+\\
&&\kappa\bigg(\int\lambda[\frac{d(r^2f^{\prime}(r))}{dr}-2(1-\frac{d(rf(r))}{dr})]dr\bigg)_{r=r_h}\bigg],\nonumber
\end{eqnarray}

\noindent and therefore, the energy changes due to the hypothetical
displacement of the horizon radius from $r_h$ to $r_h+dr_h$ is
evaluated as

\begin{eqnarray}\label{E22}
&&dE_h=E_{r_h+dr_h}-E_{r_h}=\frac{dE_h}{dr_h}dr_h=\\&&
\frac{4\pi}{\kappa}\big[dr_h+\kappa\lambda[d(r^2f^\prime(r))_{r=r_h}-2\big(dr_h-d(rf(r))_{r=r_h}\big)]\big].\nonumber
\end{eqnarray}

\noindent At the $\lambda=0$ limit, the above equations reduce to

\begin{eqnarray}\label{mse}
dE=\frac{4\pi}{2\kappa_G}\big[1-\frac{d(rf(r))}{dr}\big]dr\Rightarrow E=\frac{r}{2G}[1-f(r)],
\end{eqnarray}

\noindent nothing but the Misner-Sharp mass in the Einstein theory
\cite{ms,ref0,ref1,ref2,pad,pad1,pad2,pad3}. As a proper result, it
leads to $E_h=\frac{r_h}{2G}$ for the Misner-Sharp mass confined to
the horizon located at $r_h$. Thus, the emergence of the Einstein
result is parallel to assume $\kappa=2\kappa_G$ or equally
$\lambda=0$. It is also easy to check the $\lambda=constant$ case
recovers the result of the Rastall theory \cite{rasah}.

Now, let us use the $r-r$ component of~(\ref{gr}) to find pressure
at radius $r$ as

\begin{eqnarray}\label{p1}
&&P(r)=\frac{1}{\kappa}\big(\frac{1}{r^2}\left[r
f^\prime(r)
-1+f(r)\right]\\&&-\frac{\kappa\lambda}{r^2}[r^2f^{\prime\prime}(r)+4rf^{\prime}(r)-2+2f(r)]\big),\nonumber
\end{eqnarray}

\noindent leading to

\begin{eqnarray}\label{p2}
&&P(r_h)=\frac{1}{\kappa r_h^2}\big(r_h
f^\prime(r_h)
-1\\&&-\kappa\lambda[r_h^2f^{\prime\prime}(r_h)+4r_hf^{\prime}(r_h)-2]\big),\nonumber
\end{eqnarray}

\noindent on the event horizon where $f(r_h)=0$. Assuming $dV=4\pi
r^2 dr$ (in accordance with Eq.~(\ref{ufl10}) compatible with the
$dE=\rho dV$ relation), and using Eq.~(\ref{p2}) , we can reach

\begin{eqnarray}\label{p3}
&&P(r_h)dV=\frac{2f^\prime(r_h)}{\kappa}d(\frac{A}{4})\\&&-\frac{4\pi}{\kappa}[1+\kappa\lambda(r_h^2f^{\prime\prime}(r_h)+4r_hf^{\prime}(r_h)-2)]dr,\nonumber
\end{eqnarray}

\noindent on the event horizon ($f(r_h)=0$). Additionally, since
$f(r_h)=0$, one writes

\begin{eqnarray}\label{p4}
&&[1+\kappa\lambda(r_h^2f^{\prime\prime}(r_h)+4r_hf^{\prime}(r_h)-2)]dr\\&&
=
(1-2\kappa\lambda)dr_h+\kappa\lambda[d(r^2f^\prime(r))_{r=r_h}]+2\kappa\lambda
r_hf^\prime(r)\nonumber\\
&&=dr_h+\kappa\lambda[d(r^2f^\prime(r))_{r=r_h}-2\big(dr_h-d(rf(r))_{r=r_h}\big)],\nonumber
\end{eqnarray}

\noindent nothing but $\frac{\kappa}{4\pi}dE_h$ compared with
Eq.~(\ref{E22}). Thus, Eq.~(\ref{p3}) reduces to

\begin{eqnarray}\label{ent1}
P(r_h)dV=\frac{2f^\prime(r_h)}{\kappa}d(\frac{A}{4})-dE_h,
\end{eqnarray}

\noindent compared with the first law of thermodynamics
($PdV=TdS-dE$) \cite{ref0,ref1,ref2,pad,pad1,pad2,pad3} to get

\begin{eqnarray}\label{ent2}
TdS_h=\frac{2f^\prime(r_h)}{\kappa}d(\frac{A}{4}),
\end{eqnarray}

\noindent where $T$ and $S_h$ denote the horizon temperature and
entropy, respectively. Now, bearing the horizon temperature
($T=\frac{f^{\prime}(r_h)}{4\pi}$) in mind
\cite{ref0,ref1,ref2,pad,pad1,pad2,pad3}, the horizon entropy is
obtained as

\begin{eqnarray}\label{ent3}
dS_h=\frac{8\pi}{\kappa}d(\frac{A}{4})\Rightarrow S_h=\frac{8\pi}{\kappa}\frac{A}{4}.
\end{eqnarray}

\noindent At first sight, it looks like the Bekenstein entropy, but
in fact, it reduces to the Bekenstein entropy whenever $G$ is
constant and $\lambda=0$ parallel to the $\kappa=8\pi G=constant$
constraint. Indeed, this is the same as the entropy of the Rastall
theory, despite the fact that $\lambda=constant\neq0$ (or equally
$G$ is constant) in the Rastall theory \cite{rasah,rasen}. The
latter may be due to that $i$) both the Rastall theory and its
generalized version modify the Einstein field equations with the
same geometrical term ($R$), and $ii$) in both theories, unlike the
Rastall parameter, the Rastall gravitational coupling is constant.

%%%%%%%%%%%%%%%%%%%%%%%%%%%%%%%%%%%%%%%%%%%%%%%%%%%%%%%%%%%%%%%%%%%%%%%%%%%%%%%%%
\subsection{Entropy of the apparent horizon of FRW universe}

Now, consider a flat FRW universe with scale factor $a(t)$ and line
element

\begin{eqnarray}\label{frw}
ds^{2}=-dt^{2}+a^{2}\left( t\right) \left[dr^2+r^{2}d\Omega^{2}\right],
\end{eqnarray}

\noindent filled by an energy-momentum source specified by
$T_{\mu}^{\nu}=\textmd{diag}(-\rho,p,p,p)$. Its apparent horizon, as
the proper causal boundary
\cite{cons,cons1,cons2,cons3,Cai2,CaiKimt}, is located at
\cite{Cai2,CaiKimt}

\begin{eqnarray}\label{ah}
\tilde{r}_A=\frac{1}{H},
\end{eqnarray}

\noindent where $H=\frac{\frac{da}{dt}}{a}$ is called the Hubble
parameter, and has temperature $T=\frac{H}{2\pi}$
\cite{cons3,Cai2,CaiKimt}. During the time interval $dt$, the energy
flux crossing ($\delta Q$) the apparent horizon is evaluated as

\begin{eqnarray}\label{esv0}
\delta Q \equiv A\Psi_a dx^a,
\end{eqnarray}

\noindent in which $A$ denotes the area of the apparent horizon
\cite{Cai2,cons3}. Bearing the definition of $\Psi_a$ in mind, one
can reach

\begin{eqnarray}\label{ufl1}
\delta
Q=-\frac{3V(\rho+p)H}{2}dt+\frac{A(\rho+p)}{2}(d\tilde{r}-\tilde{r}
H dt),
\end{eqnarray}

\noindent combined with the Clausius relation ($TdS_A=\delta Q$) and
$V=\frac{4\pi}{3H^3}$ (the volume confined by the apparent horizon)
to get

\begin{eqnarray}\label{claus1}
dS_A\equiv -\frac{\delta Q}{T}=6\pi V(\rho+p)dt,
\end{eqnarray}

\noindent where $S_A$ is the entropy corresponding to the apparent
horizon (the causal boundary). Before proceeding, we use the field
equations~(\ref{gr}) for obtaining

\begin{equation}\label{1}
\frac{dH}{dt}=\dot{H}=-\frac{\kappa}{2}(\rho+p),
\end{equation}

\noindent inserted into Eq.~(\ref{claus1}) to find

\begin{eqnarray}\label{ent4}
dS_A=-\frac{12\pi V}{\kappa}dH.
\end{eqnarray}

\noindent The integral of this equation is straightforward leading
to

\begin{eqnarray}\label{ent5}
S_A=\frac{8\pi}{\kappa}\frac{A}{4},
\end{eqnarray}

\noindent for the apparent horizon entropy in the generalized
Rastall theory \cite{epjc}. The same as the static spherically
symmetric spacetimes, the obtained relation $i$) reduces to that of
the Einstein theory at the approprite limit $\kappa=8\pi G$ (or
equally $\lambda=0$), and $ii$) is the same as that of the Rastall
theory \cite{rasah,rasen}. Finally, one can use Eqs.~(\ref{claus1})
and~(\ref{1}) to see that, during the cosmic evolution, for which
$\dot{H}<0$, the second law of thermodynamics
($\frac{dS_A}{dt}\geq0$) is met whenever the $\rho+p>0$ condition is
satisfied which yields $\kappa>0$.

%%%%%%%%%%%%%%%%%%%%%%%%%%%%%%%%%%%%%%%%%%%%%%%%%%%%%%%%%%%%%%%%%%%%%%%%%%%%%%%%%%
\section{Summary}

Using the $t-t$ component of the field equations and the unified
first law of thermodynamics, we could obtain the generalized
Misner-Sharp mass, confined to the event horizon of the static
spherically symmetric spacetime with metric~(\ref{met2}), in the
generalized Rastall theory. Thereinafter, combining this result
with the $r-r$ component of the field equations, the entropy
corresponding to the horizon has been calculated. Here, we only
considered the special case for which $g_{tt}=-g^{rr}$
\cite{pad,rasah,ref2}, and the general metrics including the
$g_{tt}g_{rr}\neq constant$ cases \cite{ref2} need more
investigations considered as the future work. We also generalized
our investigation to the flat FRW universe, and got the apparent
horizon entropy. Applying the Newtonian limit to the Rastall field
equations, relation between the Newtonian gravitational coupling,
$\lambda$ and $\kappa$ has also been established which is similar
to that of the Rastall theory \cite{rastall,rasah} due to the fact
that both theories add the same geometrical term ($R$) to the
Einstein field equations.

Our study shows that the horizon entropy in the generalized Rastall
theory is the same as that of the Rastall theory \cite{rasah,rasen}
which may have two reasons including $i$) both theories modify the
Einstein theory with the same geometrical term, and $ii$)
unlike the Rastall parameter, the Rastall gravitational coupling is
constant in both theories. It is worthwhile mentioning that, the
same as the Rastall theory and also other works
\cite{rasah,rasen,plb,epjc}, the results indicate that the $S\propto
A$ relation is valid in the generalized Rastall theory, a property
preserved by the Bekenstein entropy. In both static and dynamic
cases, the obtained entropy relation reduces to the Bekenstein
entropy at the appropriate limit $\kappa=8\pi G$ (or equally
$\lambda=0$), a desired result. Finally, we saw that if $\rho+p>0$,
then the second law of thermodynamics is satisfied in this theory.

%%%%%%%%%%%%%%%%%%%%%%%%%%%%%%%%%%%%%%%%%%%%%%%%%%%%%%
\section*{Acknowledgment}
We are grateful to the anonymous referees for valuable comments.
%The work of H. Moradpour has been supported financially by
%Research Institute for Astronomy \& Astrophysics of Maragha
%(RIAAM) under research project No. 1/6025-54.
%%%%%%%%%%%%%%%%%%%%%%%%%%%%%%%%%%%%%%%%%%%%%%%%%%%%%%%

\end{document}